\documentclass[prl,twocolumn,superscriptaddress,nofootinbib]{revtex4-1}
\usepackage{graphicx}
\usepackage{bm}
\usepackage{latexsym,amssymb,amsmath,amsfonts,amssymb,txfonts,wasysym,float}
\usepackage{mathrsfs}

\usepackage{enumitem}
\usepackage{slashed}
\usepackage{comment}

\usepackage{xcolor}

\usepackage{textcomp}

\usepackage{array, caption,  tabularx, makecell, booktabs}%
\usepackage[font=footnotesize]{caption}

\newcommand{\sigmabar}{\bar{\sigma}}

\newcommand{\ep}{\epsilon}

\newcommand{\ept}{\tilde{\epsilon}}
\DeclareMathOperator\arctanh{arctanh}

\newcommand{\psib}{\bar{\psi}}

\newcommand{\gammab}{\bar{\gamma}}

\newcommand{\lambdab}{\bar{\lambda}}

\newcommand{\im}{\mathrm{i}}
\newcommand{\esi}{\left(\epsilon\cdot\sigma\right)}
\newcommand{\esib}{\left(\epsilon\cdot\bar{\sigma}\right)}

\begin{document}
\title{ Open Superstring First Mass Level Effective Lagrangian: \\ Massive Spin-$3/2$ Fields in an Electromagnetic Background}

\author{Karim Benakli}

\email{kbenakli@lpthe.jussieu.fr}

\affiliation{Sorbonne Universit\'e, CNRS,\\ Laboratoire de Physique Th\'eorique et Hautes Energies, LPTHE, F-75005 Paris, France.
}

\author{Cassiano A. Daniel}
\email{c.daniel@unesp.br}
\affiliation{ICTP South American Institute for Fundamental Research \\
Instituto de F\'{i}sica Te\'{o}rica, Universidade Estadual Paulista \\
Rua Dr. Bento Teobaldo Ferraz 271, 01140-070, S\~{a}o Paulo - SP, Brasil.
}

\author{Wenqi Ke}

\email{wke@lpthe.jussieu.fr}
\affiliation{Sorbonne Universit\'e, CNRS,\\ Laboratoire de Physique Th\'eorique et Hautes Energies, LPTHE, F-75005 Paris, France.
}

\begin{abstract}
 \vskip 2mm \noindent We derive fully explicit equations of motion, and the associated set of constraints, describing the propagation in a flat space-time of a charged spin-$3/2$ massive state in a constant electromagnetic background.  For this purpose, we provide the Lagrangian for the physical fermionic fields in the first massive level of the open superstring. We first write a compact Lagrangian, allowing a simple derivation of the equations of motion and constraints. Then another one is given that yields directly a decoupled system of equations, though the fields of different spins look coupled at the level of the Lagrangian.
\end{abstract}

\maketitle

\section{INTRODUCTION}

The description of the propagation of massive spin-3/2  charged states in an electromagnetic background is an old problem dating back to the 1930's \cite{Dirac:1936tg,Fierz:1939ix}. An important obstacle is the appearance of superluminal propagation which leads to a loss of causality \cite{Velo:1969bt}. The difficulty of writing causal equations of motion and a Lagrangian is illustrated, for example, in the work of \cite{Deser:2000dz} where it has been shown that a large class of modifications of the minimal theory fail to restore causality. However, a solution to the problem is in principle possible. Indeed, in \cite{Porrati:2009bs}, an ansatz for a Lagrangian has been proposed whose coefficients can be obtained recursively, order by order in the field strength $F_{mn}$ of the electromagnetic background. In a previous work \cite{Benakli:2021jxs}, we generalised to the charged case the computation in \cite{Berkovits:1998ua,Berkovits:1997zd} of the superspace Lagrangian and the equations of motion of the first massive level of the open superstring. We have thus generalised the work of \cite{Argyres:1989cu,Porrati:2010hm,Klishevich:1998sr} for the bosonic case. Here we will obtain explicit expressions for the Lagrangian of fermionic physical states and their equations of motion. 
The details of the derivation of our results will be given elsewhere \cite{Toappear}.

The first massive level of open superstrings has  on-shell 12 complex fermionic degrees of freedom of mass $M$. Eight of them correspond to massive spin-3/2 fields denoted $(\lambda_{mj}, \chi_{mj})$ and four to spin-1/2  ones $\psi_{j}, \gamma_{j}$, with $j=1,2$. They carry total charges $Q = q_0 + q_\pi$ where  $q_0$ and $q_\pi$ are the two charges at the string end-points. The covariant derivative  is
\begin{equation}\begin{aligned}
  D_n =  (\partial -i Q F\cdot X)_n
    \end{aligned}\label{D_n}
\end{equation}
where $X^n$ is a space-time coordinate. The stringy origin of these states manifests itself in the dependence on $F_{mn}$ through the matrix \cite{Fradkin:1985qd,Abouelsaood:1986gd}
\begin{equation}\begin{aligned}
  \epsilon =  \frac{\Lambda^2}{\pi} \left[\arctanh\left(\frac{\pi q_0 F}{\Lambda^2}\right) +\arctanh\left(\frac{\pi q_\pi F}{\Lambda^2}\right)\right]
    \end{aligned}\label{epsilon}
\end{equation}
where $\Lambda$ is the fundamental (string) scale of the theory and the dressed form of the covariant derivative \cite{Abouelsaood:1986gd,Argyres:1989cu}
\begin{equation}\begin{aligned}
\mathfrak{D}_m =  -\im \,  \mathfrak{M}_{mn}\,  D^n ,  \qquad \left[\mathfrak{D}_m , \mathfrak{D}_n \right] =\im \epsilon_{mn}
    \end{aligned}\label{mathfrakD_n}
\end{equation}
 with the matrix $\mathfrak{M}$ satisfying:
\begin{equation}\begin{aligned}
  \mathfrak{M} \cdot \mathfrak{M}^T= \frac{\epsilon}{Q F}
    \end{aligned}\label{Mmatrix}
\end{equation}
As noted in \cite{Argyres:1989cu} for the case of the bosonic open string, the consistency of the Lagrangian and the derivation of the equations of motion use the antisymmetric property of $\epsilon_{mn}$ but nowhere the explicit dependence of $\epsilon_{mn}$ on $F_{mn}$. Therefore, our analysis continues to hold if we take everywhere the quantum field theory limit $\epsilon_{mn} \rightarrow Q F_{mn}$ and $\mathfrak{D}_m \rightarrow D_m$.

\section{A Compact Fermionic Lagrangian}

 For simplicity, we define the dual tensor $\ept_{mn}\equiv\varepsilon_{mnkl}\ep^{kl}/2$, and the contractions  $\esi\equiv \ep_{mn}\sigma^{mn}$, $\esib\equiv\ep_{mn}\sigmabar^{mn}$, $\left(\ep\ep\right)\equiv\ep^{mn}\ep_{mn}$, $\left(\ep\ept\right)\equiv\ep^{mn}\ept_{mn}$. The bold symbols, \textit{e.g.}~$\boldsymbol{\lambdab}^m$, include a rescaling factor $\boldsymbol{\lambdab}^m\equiv\left(\eta_{mn}-\im\frac{2}{M^2}\ep_{mn}\right)\lambdab^n$. Expanding the superspace action in \cite{Benakli:2021jxs} in components, and eliminating unphysical fields, we obtain:
\begin{equation}
\begin{aligned}
    \mathcal{L}_F=&-\frac{\mathrm{i}}{\sqrt{2}}\left[2\left( \lambda_1^m \sigma^n\mathfrak{D}_n\bar{\boldsymbol{\lambda}}_{1m}
\right)  +\left(\bar{\boldsymbol\chi}_{1m} \bar{\sigma}^n{\sigma}^k\bar{{\sigma}}^m\mathfrak{D}_k\boldsymbol\chi_{1n}\right)\right]
\\&-\sqrt{2}M \left[ \left({\boldsymbol{\lambda}}_1^m{\boldsymbol{\chi}}_{1m} \right)+\text{h.c.}\right] 
\\&+\sqrt{2}\left[-\frac{\im}{4}\left(\psi_1\sigma^m \mathfrak{D}_m\psib_1 \right)+2\mathrm{i}\left( \gamma_1\sigma^m\mathfrak{D}_m\gammab_1\right)\right]
\\&+ \left[3\left({\boldsymbol{\chi}}_1^m \sigma_{mn}\mathfrak{D}^n \psi_1\right)-\frac{1}{2}\left(\boldsymbol{\chi}_1^m\mathfrak{D}_m\psi_1 \right)
-2\left(\boldsymbol{\lambda}_1^m\mathfrak{D}_m \gamma_1\right)
\right.
\\& \left.\quad  -\frac{\im}{2}M \left( \boldsymbol{\lambda}_1^m\sigma_m\psib_1\right)-2\mathrm{i}M\left(\bar{\boldsymbol{\chi}}_{1}^m\sigmabar_m\gamma_1\right)+\text{h.c.} \right]
\\&+M\left[\frac{1}{\sqrt{2}}\left( \psi_1\gamma_1\right)+\text{h.c.}\right]+\left(1\leftrightarrow 2 \right)
\\& -\frac{1}{M}\left[\bar{\boldsymbol{\chi}}_1^m \left(\epsilon\cdot\sigmabar \right)\sigmabar_m\gamma_1+{\boldsymbol{\chi}}_2^m\left( \epsilon\cdot\sigma\right)\sigma_m \gammab_2+\text{h.c.}\right]
\end{aligned}\label{chi-redefined}
\end{equation}

which leads to the coupled equations of motion:
\begin{equation}
\begin{aligned}\im &\left(\sigma^n\mathfrak{D}_n\boldsymbol{\bar{\lambda}}_{1m}\right)_\alpha=-M\boldsymbol{\chi}_{1m\alpha}+\im  \frac{2}{M}\epsilon_{mn} \boldsymbol{\chi}_{1\alpha}^n -\frac{M}{2\sqrt{2}}\im\left(\sigma_m\psib_1\right)_\alpha 
\\&  \quad  -\frac{1}{\sqrt{2}M}\epsilon_{mn}\left(\sigma^n\psib_1\right)_\alpha 
-{\sqrt{2}}\mathfrak{D}_m\gamma_{1\alpha}+\im\frac{2\sqrt{2}}{M^2}\epsilon_{mn}\mathfrak{D}^n\gamma_{1\alpha}
\label{eqlam}
\end{aligned}
\end{equation}

\begin{equation}
\begin{aligned}
&\mathrm{i}\sqrt{2}\left(\sigmabar^n  \sigma^k\sigmabar_m\mathfrak{D}_k\boldsymbol{\chi}_{1n} \right)^{\dot{\alpha}}=-2\sqrt{2}M\boldsymbol{\lambdab}_{1m}^{\dot{\alpha}}+6\left(\sigmabar_{mn}\mathfrak{D}^n\psib_1\right)^{\dot{\alpha}}  
\\& \quad
  -\mathfrak{D}_m\psib_1^{\dot{\alpha}}-4\mathrm{i}M\left(\sigmabar_m\gamma_1\right)^{\dot{\alpha}} -\frac{2}{M}\left[\left(\epsilon\cdot\sigmabar \right)\sigmabar_m\gamma_1\right]^{\dot{\alpha}}
\end{aligned}  \label{eqchi}
\end{equation}

\begin{equation}
\begin{aligned}
    &\im\left(\sigma^m\mathfrak{D}_m\psib_1\right)_\alpha=-6\sqrt{2}\left(\sigma_{mn}\mathfrak{D}^m\boldsymbol{\chi}_1^n\right)_\alpha+{\sqrt{2}}\mathfrak{D}^m\boldsymbol{\chi}_{1m\alpha}  
    \\& \qquad \qquad  \qquad  \qquad
    +\sqrt{2}M\im\left(\sigma^m\boldsymbol{\lambdab}_{1m} \right)_\alpha+2M\gamma_{1\alpha}
    \label{eqpsi}
\end{aligned}
\end{equation}

\begin{equation}
\begin{aligned}    
& \mathrm{i}\left(\sigmabar^m\mathfrak{D}_m\gamma_1 \right)^{\dot{\alpha}}=-\mathrm{i}\frac{M}{\sqrt{2}}\left(\sigmabar^m\boldsymbol{\chi}_{1m}\right)^{\dot{\alpha}}-\frac{\sqrt{2}}{2}\mathfrak{D}^m\boldsymbol{\lambdab}_{1m}^{\dot{\alpha}}  
        \\& \qquad \qquad \qquad
        -\frac{1}{2\sqrt{2} M}\left[\sigmabar^m\left(\epsilon\cdot\sigma \right)\boldsymbol{\chi}_{1m} \right]^{\dot{\alpha}} -\frac{M}{4}\psib_{1}^{\dot{\alpha}}\label{eqgamma}
       \end{aligned}
\end{equation}

In absence of the electromagnetic background, they give
\begin{equation}
    \begin{aligned}
    &\im\left(\sigma^m\partial_m\psib\right)_\alpha=-M\gamma_{\alpha},\quad \im\left(\sigmabar^m\partial_m\gamma \right)^{\dot{\alpha}}=-M\psib^{\dot{\alpha}}
    \\&
    \left(\sigmabar^m\chi_m\right)^{\dot{\alpha}}=0,\quad \left(\sigma^m\bar{\lambda}_m\right)_\alpha=\frac{3}{\sqrt{2}}\mathrm{i}\gamma_\alpha,
    \\& \quad\partial^m\chi_{m\alpha}=0,\quad\partial^m\lambdab_m^{\dot{\alpha}}=\frac{3M}{2\sqrt{2}}\psib^{\dot{\alpha}}
    \end{aligned}
\end{equation}

The  spin-$3/2$  and spin-1/2 fields can be decoupled in these equations by performing the field redefinition: 
\begin{equation}
\lambdab_m^{\prime\dot{\alpha}}=\lambdab_m^{\dot{\alpha}}+\frac{\mathrm{i}}{2\sqrt{2}}\left(\sigmabar_m\gamma\right)^{\dot{\alpha}}-\frac{1}{\sqrt{2}M}\partial_m\psib^{\dot{\alpha}}\label{modiflam}\end{equation}
but, in the presence of an electromagnetic background, the redefinition of the fields is more complicated. To obtain a similar form of the equations of motion, in this case, some tedious, but straightforward, algebraic manipulations are necessary to derive first the constraints. Combining the $\sigma$-trace and divergence of the equations of motion, as well as applying $\sigma^m\esib $ on them, leads to:
\begin{equation}
\begin{aligned}    
\sigmabar^m\boldsymbol{\chi}_{1m}=&0
\\
\mathfrak{D}^m\boldsymbol{\chi}_{1m}=&-\frac{\im }{\sqrt{2}M}\esi\gamma_1
\\
\sigma^m\boldsymbol{\lambdab}_{1m}=&\frac{3}{\sqrt{2}}\im\gamma_1-\frac{\sqrt{2}}{M^2}\esi\gamma_1
\\
    \mathfrak{D}^m\boldsymbol{\lambdab}_{1m}=&\frac{3M}{2\sqrt{2}}\psib_1-\frac{1}{2M}\sigmabar^m\esi\boldsymbol{\chi}_{1m}
\end{aligned}
\end{equation}
in addition to the Dirac equations of $\psib_1$ and  $\gamma_1$:
\begin{equation}
   \im\sigma^m\mathfrak{D}_m\psib_1=-M\gamma_{1},\quad   \im\sigmabar^m\mathfrak{D}_m\gamma_1=-M\psib_{1}
\end{equation}
Similar coupled constraints are also obtained in  \cite{Benakli:2021jxs}. However, we can go further and look for a generalisation of \eqref{modiflam} to decouple the spin-3/2 and spin-1/2 on shell. The new spin-3/2 definition reads:
\begin{equation}
\begin{aligned}     \boldsymbol\lambdab^{\prime}_{1m}&\equiv\boldsymbol\lambdab_{1m}+\frac{\im}{2\sqrt{2}}\left[1-\im \frac{2  }{M^2}\esib\right]\sigmabar_m\gamma_1
\\& \quad -\frac{1}{\sqrt{2}M} \left[\eta_{mn}
-\im  \frac{2  }{M^2}\left(\epsilon_{mn}+\im \tilde{\epsilon}_{mn}\right)\right]\mathfrak{D}^n\psib_1
 \\\boldsymbol{\chi}_{1m}^\prime&\equiv \boldsymbol{\chi}_{1m}+\frac{1}{\sqrt{2} M^2}\esi\sigma_m\psib_1
        \end{aligned}\label{newsp32}
\end{equation}
leading to the equations of motion and constraints:
\begin{equation}
\begin{aligned}   
&\im\sigmabar^m\mathfrak{D}_m\gamma_1=-M\psib_1,\quad \im \sigma^m\mathfrak{D}_m\psib_1=-M\gamma_{1}
\\&  \mathrm{i}\sigmabar^n\mathfrak{D}_n\boldsymbol{\chi}^\prime_{1m}  =-M\boldsymbol{\lambdab}^{\prime}_{1m},
\\&
\mathrm{i}\sigma^n\mathfrak{D}_n \boldsymbol{\lambdab}^{\prime}_{1m}=-M\left(\eta_{mn}-\im  \frac{2}{M^2} \epsilon_{mn}\right)\boldsymbol{\chi}^{\prime n\ }_{1}\\&\mathfrak{D}^m\boldsymbol{\chi}^\prime_{1m}=0,\quad  \sigmabar^m\boldsymbol{\chi}^\prime_{1m}=0\\&\mathfrak{D}^m\boldsymbol{\lambdab}_{1m}^{\prime} =-\frac{1}{2 M}\sigmabar^m\esi\boldsymbol{\chi}_{1m}^\prime,\quad \sigma^m\boldsymbol{\lambdab}_{1m}^{\prime}=0
 \end{aligned}  \label{twocomp-onshell}
\end{equation}
which trivially give the well-known Dirac and Fierz-Pauli system derived from the Rarita-Schwinger Lagrangian for $\epsilon=0$.

\subsection{Four-component notation:}

Let's express the spin-1/2 and spin-3/2 in four components, introducing 
\begin{equation}
    \Phi_1\equiv \begin{pmatrix} \gamma_{1\alpha} \\
\bar{\psi}_1^{\dot{\alpha}}
\end{pmatrix},\quad \boldsymbol{\Psi}_{1m}\equiv \begin{pmatrix}\boldsymbol{\chi}^\prime_{1m\alpha} \\
\bar{\boldsymbol{\lambda}}_{1m}^{\prime\dot{\alpha}}
\end{pmatrix} 	
\end{equation}
and the $\gamma$-matrices 
\begin{equation}
      \gamma^m=\begin{pmatrix}0&\sigma^m \\\sigmabar^m&0
\end{pmatrix},\quad \gamma^5=\gamma^0\gamma^1\gamma^2\gamma^3=\begin{pmatrix}-\im&0\\0&\im
\end{pmatrix}\end{equation}
For shorthand, we note $\slashed{\mathfrak{D}}\equiv \gamma^m\mathfrak{D}_m
$. The spin-1/2 satisfies the Dirac equation in QED:
\begin{equation}
\left(\im 
\slashed{\mathfrak{D}}+M\right)\Phi_1=0
\end{equation}

As for spin-3/2, the constraints can be written as 
\begin{equation}  
\left[\mathfrak{D}^m-\frac{1}{2 M}\left(\epsilon^{mn}
+\im\tilde{\epsilon}^{mn}\right)\gamma_n\right]\boldsymbol{\Psi}_{1m}=0  ,
\label{constraint-derivative}
\end{equation}
\begin{equation}
\gamma^m\boldsymbol{\Psi}_{1m}=0
\label{constraint-trace}
\end{equation}

with equation of motion
\begin{equation}
\left(\im 
\slashed{\mathfrak{D}}+M\right)\boldsymbol{\Psi}_{1m}=\im  \frac{2}{ M}\left( \epsilon_{mn}\boldsymbol{\Psi}_{1L}^n    \right)
\end{equation}

The projection operator is given by  $P_L=(1+\im \gamma^5)/2$. The fermions with index 2 correspond to the conjugates of the index 1 in the free case. Their equations of motion and constraints can be worked out in an analogous way. The results correspond to the replacements: $1\leftrightarrow 2 $ and $\ep\leftrightarrow-\ep$.

\section{Spinors with decoupled equations of motion}

We can perform the last field redefinition \eqref{newsp32} at the level of the Lagrangian \eqref{chi-redefined} directly. For this purpose, it is useful to denote the inverse matrix by $G_{mn}\equiv\left(\eta_{mn}-\im  \frac{2}{ M^2}\epsilon_{mn}\right)^{-1}$. For the index 1, the redefinitions  are the following:\begin{equation}
    \begin{aligned}       \boldsymbol\lambdab_{1m}\rightarrow&\boldsymbol\lambdab_{1m}-\frac{\im}{2\sqrt{2}}\left[1- \im\frac{2}{ M^2}\esib\right]\sigmabar_m\gamma_1\\&\qquad +\frac{1}{\sqrt{2}M} \left[\eta_{mn}-\im \frac{2}{ M^2} \left(\epsilon_{mn}+\im\tilde{\epsilon}_{mn}\right)\right]\mathfrak{D}^n\psib_1
 \\\boldsymbol{\chi}_{1m}\rightarrow& \boldsymbol{\chi}_{1m}-\frac{1}{\sqrt{2}M^2}\esi\sigma_m\psib_1
   \end{aligned}\label{redef-comp}
\end{equation}
Those with index 2 correspond simply to \eqref{redef-comp} with $1\leftrightarrow2$, $\ep\leftrightarrow -\ep$. The resulting Lagrangian:

\begin{equation}
\begin{aligned}
    \mathcal{L}_F=&-\im\sqrt{2}\left[\left( \lambda_1^m \sigma^n\mathfrak{D}_n\bar{\boldsymbol{\lambda}}_{1m}
\right)  +\frac{1}{2}\left(\bar{\boldsymbol\chi}_{1m} \bar{\sigma}^n{\sigma}^k\bar{{\sigma}}^m\mathfrak{D}_k\boldsymbol\chi_{1n}\right)\right]
\\&-\sqrt{2}M\left[ \left({\boldsymbol{\lambda}}_1^m{\boldsymbol{\chi}}_{1m} \right)+\text{h.c.}\right]+\frac{\sqrt{2}}{M}\left[\gammab_1\mathfrak{D}^2\psib_1+\text{h.c.}\right]
\\&+ \left[\frac{3}{2}\left(\boldsymbol{\chi}_1^m\sigma_m\sigma_n\mathfrak{D}^n\psi_1 \right)-\frac{\im M}{2} \left( \boldsymbol{\lambda}_1^m\sigma_m\psib_1\right)\right.
\\&\left.\quad -\frac{3 M}{2}\mathrm{i}\left(\bar{\boldsymbol{\chi}}_{1}^m\sigmabar_m\gamma_1\right)-2\left(\boldsymbol{\lambda}_1^m\mathfrak{D}_m \gamma_1\right)+\text{h.c.} \right] 
\\&+\sqrt{2}\left[\frac{\im}{4}\left(\psi_1\sigma^m \mathfrak{D}_m\psib_1 \right)+\mathrm{i}\left( \gamma_1\sigma^m\mathfrak{D}_m\gammab_1\right)\right]
\\&-\frac{\im\sqrt{2}}{8M^4}G^{mn}\gammab_1\sigmabar_m\left[M^2-2\im\esi\right]\sigma_k
\\&
\qquad \times  \left[M^2-2\im  \esib\right]\sigmabar_n\mathfrak{D}^k\gamma_1
\\&+\frac{\im}{\sqrt{2}M^2}G^{mn}\left[\eta_{mp}+\im\frac{2}{ M^2}\left(\epsilon_{mp}-\im \tilde{\epsilon}_{mp}\right)\right]\times
\\&\quad \left[\eta_{nq}-\im \frac{2}{ M^2}\left(\epsilon_{nq}+\im\tilde{\epsilon}_{nq}\right)\right]\psi_1\sigma^k\mathfrak{D}^p\mathfrak{D}_k\mathfrak{D}^q\psib_1
\\&-\frac{\im\sqrt{2}}{M^4}\left(\epsilon_{mn}\epsilon^{mk}+\tilde{\epsilon}_{mn}\tilde{\epsilon}^{mk}\right)\psi_1\sigma_k\mathfrak{D}^n\psib_1
\\& -{\im}\frac {\sqrt{2}} { M^2}\tilde{\epsilon}_{mn}\left(\psi_1\sigma^n\mathfrak{D}^m\psib_1\right)
+\im \frac {2 \sqrt{2}} { M^2}\tilde{\epsilon}_{mn}\left(\gammab_1\sigmabar^n\mathfrak{D}^m\gamma_1\right)
\\& 
+\frac{1}{2\sqrt{2}M}\left\{G^{mn}\left[\eta_{kp}-\im \frac {2} { M^2} \left(\epsilon_{kp}+\im\tilde{\epsilon}_{kp}\right)\right]\times  
\right.
\\&\left.\qquad \quad \left[\eta_{nl}-\im \frac {2} { M^2} \left(\epsilon_{nl}+\im\tilde{\epsilon}_{nl}\right)\right]\gammab_1\sigmabar_m\sigma^p\mathfrak{D}^k\mathfrak{D}^l\psib_1+\text{h.c.}\right\}
\\&-\left\{\frac{\im}{M}\left[\eta_{mk}-\im \frac {2} { M^2} \left(\epsilon_{mk}+\im\tilde{\epsilon}_{mk}\right)\right]\lambda_1^m\sigma_n\mathfrak{D}^n\mathfrak{D}^k\psib_1\right.
\\&\left.\quad +\frac{1}{2}\lambda_1^m\sigma_n\left[1-\im \frac {2} { M^2} \esib\right]\sigmabar_m\mathfrak{D}^n\gamma_1+\text{h.c.}\right\}
\\&
+\frac {1} { M}\left[\boldsymbol{\lambda}_1^m\esi\sigma_m\psib_1+\frac {\sqrt{2}} { M^2}\left(\epsilon\epsilon-\im \epsilon\tilde{\epsilon}\right)\psi_1\gamma_1+\text{h.c.}\right]\\&+\left(1\leftrightarrow 2, \epsilon\leftrightarrow-\epsilon \right)
\end{aligned}\label{comp-redefined}
\end{equation}
gives decoupled equations of motion and constraints in \eqref{twocomp-onshell}, without the need to redefine the fields. Still, the Lagrangian \eqref{chi-redefined} has the advantage of not including  higher derivatives.

\section{CONCLUSIONS}


One also expects to have a Lagrangian of the fermionic fields in the first massive level of the open superstring that reduces, in absence of an electromagnetic background, to a sum of the usual Rarita-Schwinger and Dirac Lagrangians. We have indeed obtained such a Lagrangian, but we have chosen not to present it here because it is a long expression that is not very enlightening. It contains three parts: the first one is the sum of Rarita-Schwinger and Dirac Lagrangians, up to replacing the covariant derivatives by partial derivatives, whereas the others represent additional kinetic plus the mass terms, and  couplings between spin-$3/2$ and spin-$1/2$. The last two pieces disappear with vanishing electromagnetic field. This will be provided in \cite{Toappear}.

We have found fully explicit equations of motion and constraints for the massive charged spin-3/2 field in a constant electromagnetic background. It is interesting to note that in the case of \cite{Deser:2000dz} the $\gamma$-trace constraint is modified. As a result, it disappears for a particular value of the electromagnetic field strength. This signals  
the presence of the Velo-Zwanziger problem. Here, \eqref{constraint-trace} is the same form as in the case without electromagnetic background and is always present, as it is in the proposal of \cite{Porrati:2009bs}. The Lagrangian as derived by string field theory in \cite{Benakli:2021jxs} involves the presence of spin-1/2 fields to enforce the constraints. 

\section{ACKNOWLEDGEMENTS}

We are grateful to Nathan Berkovits for useful discussions.
CAD acknowledges FAPESP grant numbers 2020/10183-9 and 2022/14599-0 for financial
support. The work of WK is
supported by the Contrat Doctoral Sp\'ecifique Normalien (CDSN) of Ecole Normale Sup\'erieure –
PSL.

\bibliographystyle{apsrev4-1}
\bibliography{refs} 

\end{document}